\documentclass{article}
\usepackage{spconf,amsmath,amssymb,graphicx,hyperref,booktabs,array,microtype,xcolor}

\graphicspath{{figures/}}
\usepackage[most]{tcolorbox}
\usepackage{graphicx}  
\usepackage{caption}   
\usepackage{float} 
\usepackage[scaled=0.85]{beramono}
\usepackage{booktabs}
\usepackage{tabularx}
\usepackage{wrapfig}   
\newcolumntype{Y}{>{\raggedright\arraybackslash}X}
\usepackage[table]{xcolor}   
\definecolor{vGreen}{HTML}{E6F4EA}   
\definecolor{vBlue}{HTML}{E3EEFB}    
\definecolor{vRose}{HTML}{FBE7E7}    
\definecolor{tGreen}{HTML}{2E7D32}
\definecolor{tBlue}{HTML}{1F5FA8}
\definecolor{tRose}{HTML}{B3261E}
\definecolor{hdr}{HTML}{F2F2F2}

\definecolor{iitm}{HTML}{8B1A1A}     
\definecolor{illini}{HTML}{FF5F05}   
\definecolor{indep}{HTML}{555555}    
\definecolor{amazon}{HTML}{FF9900}   

\definecolor{google}{HTML}{4285F4}

\newtcolorbox{findingbox}{
  enhanced,
  colback=vBlue!60,
  colframe=tBlue,
  boxrule=0pt,
  leftrule=2.5pt,
  arc=2pt,
  sharp corners=west,
  left=6pt, right=6pt, top=4pt, bottom=4pt,
  fontupper=\small,
  before skip=6pt, after skip=6pt
}

\newcommand{\finding}[2]{%
  \begin{findingbox}
  \textcolor{tBlue}{\textbf{Finding #1.}}\ #2
  \end{findingbox}}

\IfFileExists{ptmr7t.tfm}{}{%
  \typeout{** NOTE: PostScript fonts missing, using Computer Modern **}}

\def\Ci#1{{\small[#1]}}
\newcommand{\Fx}{\ensuremath{\overline{\mathrm{F1}}}}
\newcommand{\Fxb}{\ensuremath{\overline{\mathbf{F1}}}}

\title{Derivation Depth Exposes Reasoning Failures in Tool-Augmented Music Analysis}

\makeatletter
\def\name#1{\gdef\@name{{\small\em #1}}}
\makeatother

\name{Rupali Rajesh$^{1*}$, Sethuraman T V$^{2*}$, Savya Khosla$^{2}$, Srinidhi Sunkara$^{5}$, Aditya Shanmugham$^{4}$, Rohan Maheshwari$^{3}$, Derek Hoiem$^{2}$}

\address{\footnotesize
\textcolor{iitm}{$^{1}$IIT Madras} \quad
\textcolor{illini}{$^{2}$UIUC} \quad
\textcolor{indep}{$^{3}$Independent} \quad
\textcolor{amazon}{$^{4}$Amazon Science} \quad
\textcolor{google}{$^{5}$Google} \quad
$^{*}$Equal contribution}

\begin{document}
\ninept
\raggedbottom
\setlength{\textfloatsep}{10pt plus 2pt minus 2pt}
\setlength{\floatsep}{8pt plus 2pt minus 2pt}
\maketitle



\begin{abstract}
Audio-language models remain unreliable at extracting basic musical properties from audio, motivating pipelines that place a music information retrieval (MIR) tool between the audio and the model. The tool reports measurements such as keys, chords, and structural similarity, and the model is expected to derive the answer. We test whether it can. Holding the musical information fixed, we vary derivation depth: the number of reasoning steps between the tool output and the answer. When the report states the conclusion, all five models score near $100\%$, and they follow it even when it is wrong. One step away, accuracy drops. On modulation, open models fall to $57$--$81\%$, although they answer that single step correctly when asked directly, and reach $0\%$ at greater depths. Frontier models degrade less, but still degrade. Adding the audio does not recover the loss. The bottleneck in tool-augmented music analysis is not access to musical information but reasoning from it, so tool design must account for how much derivation it leaves to the model. \href{https://sethuramanio.github.io/tool_augmented_music_understanding/}{Project Page}
\end{abstract}

\section{Introduction}
\label{sec:intro}

Audio-language models offer a natural interface for music analysis, but achieve only near-chance performance when recovering basic musical properties directly from audio \cite{muse,mmau}. A promising alternative
is to augment them with music information retrieval (MIR) tools. Rather than requiring the language model to recognize every property from audio directly, an MIR system can provide keys, chords, structural similarity, or other measurements, which the model then uses to answer higher-level musical questions. Classical MIR methods already provide many such measurements
\cite{foote,krumhansl,librosa}, and recent tool augmented audio systems report substantial gains from incorporating external evidence
\cite{echoagent,msef}.

However, MIR tools often return intermediate measurements rather than the answer itself. For example, a tool may return ``new key: D major'', while the user's question asks whether the modulation is to a major or
minor key: even this requires the language model to recognize that D major is a major key.  Existing tool-augmented evaluations measure only final accuracy, which conflates tool errors with failures to reason over tool output. We isolate the second: given the right musical information, can the model derive the answer. 

To isolate this capability, we introduce \textbf{derivation depth}: the number of inferential steps needed to get from a tool's measurement to
the answer, ranging from zero, where the report already states the conclusion, to several, where it gives only a raw measurement.

\begin{figure}[H]
    \centering
    \captionsetup{font=small}
    \includegraphics[width=0.4\textwidth,trim=26 61 26 61,clip=true]{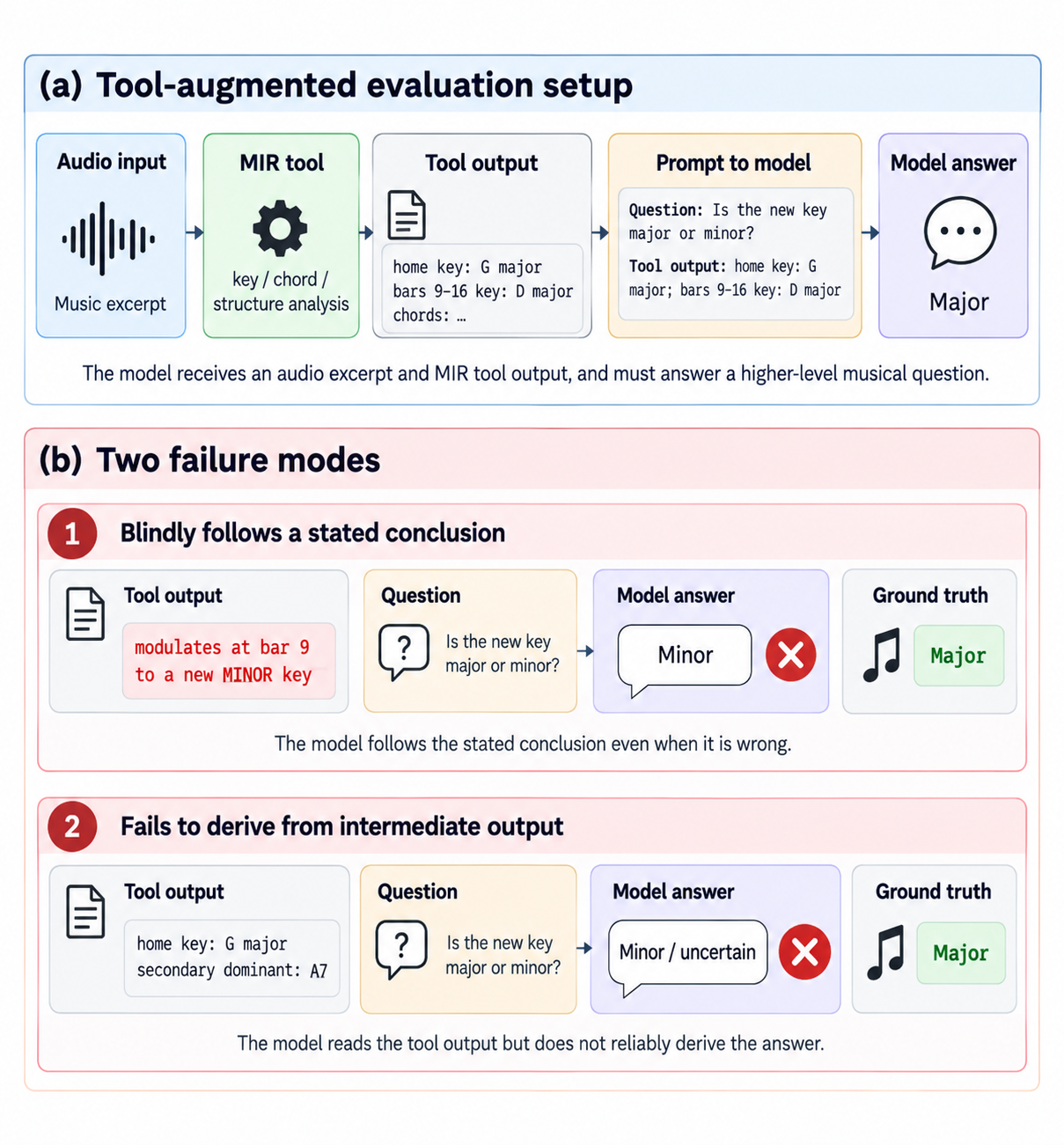}
    \caption{\textbf{Setup and failure modes.} (a) The model answers a musical question from the audio and an MIR tool's output. (b) It follows a stated conclusion even when wrong (top), and fails to derive the conclusion from intermediate output such as the home key and secondary dominant (bottom).}
    \label{fig:teaser}
\end{figure}

\noindent Fig.~\ref{fig:teaser} shows the setup and the two failures we measure. We evaluate five audio-language models across five music-analysis tasks, using controlled stimuli and real recordings. We additionally test whether the original audio can compensate for failed derivation, probe the musical knowledge required at each depth, and deliberately invert tool conclusions to measure whether models verify or simply follow them.

The results expose a reliability gap. When the tool states the conclusion directly, models are nearly perfect; when that conclusion is deliberately inverted, they follow the incorrect statement just as readily. More importantly, even one elementary reasoning step is enough to substantially reduce accuracy. On modulation, replacing ``the new key is major'' with ``new key: D major'' drops the three open models from $100\%$ to
$57$--$81\%$, despite all three identifying the same major/minor fact with $100\%$ accuracy when asked directly. Providing the chord instead drops accuracy to $0$--$26\%$, and providing only its pitches reduces it to $0\%$. Thus, the difficulty arises before long reasoning chains are needed: models can possess the required musical knowledge yet fail to apply it to
the tool output.

The effect persists with acccess to audio and on real GuitarSet recordings. For modulation, the answer never becomes decodable inside the network. For structure, it emerges mid-network but fades before the output. This limits a natural workaround: writing out an intermediate step helps when the network forms the information, but does not rescue deeper derivations and can discard what the final answer needs.

Together, these results identify a missing capability in tool-augmented music analysis: \textcolor{tBlue}{reliable reasoning over tool outputs}. Improving
the MIR tool alone is insufficient when the downstream model cannot reliably transform its measurements into the requested conclusion. In summary, our
contributions are a \textcolor{tBlue}{controlled framework for measuring derivation depth}, evidence that \textcolor{tBlue}{even elementary
derivations fail despite available musical knowledge}, and mechanistic
evidence distinguishing conclusions that are
\textcolor{tBlue}{never formed} from those that are
\textcolor{tBlue}{formed but lost before output}.


\section{Method and Experiments}
\label{sec:method}
Experiments fix the musical fact and varies only how it reaches the model: report format, audio, and reasoning distance.
\subsection{Setup}
\label{sec:stimuli}

\noindent\textbf{Tasks.} Each task is a minimal pair: two versions of a piece differing in exactly one musical property, which isolates that property as the sole determinant of the correct answer. The five tasks are cadence (authentic vs.\ deceptive), modulation (new major vs.\ minor key), scale (Lydian vs.\ Ionian, one note apart), motif (transposed return vs.\ a new motif) and structure (ABA vs.\ through-composed). Each targets a property a standard MIR tool reports. All battery items are synthetic: short scores written to exhibit exactly one property are generated programmatically as MIDI and rendered to audio with FluidSynth (TimGM6mb General MIDI soundfont, piano), and the template report is filled from the known score, so ground truth is exact. Each task has $N{=}6$ compositions rendered at up to five transpositions between $-6$ and $+6$ semitones, giving up to $30$ pairs per task ($23$ for scale).

\begin{table}[t]
\centering
\caption{One modulation item rendered in each tool-output format. Colors group formats: verdict (\textcolor{tGreen}{green}), value or prose (\textcolor{tBlue}{blue}), raw tool output (\textcolor{tRose}{rose}).}
\label{tab:formats}
\footnotesize
\setlength{\tabcolsep}{3pt}
\renewcommand{\arraystretch}{1.0}
\begin{tabularx}{\columnwidth}{@{}l l >{\raggedright\arraybackslash}X@{}}
\toprule
\rowcolor{hdr}
 & Style & Example tool output \\
\midrule
\rowcolor{vGreen}
\textcolor{tGreen}{\textbf{F1}}  & Verdict        & ``The piece modulates at bar 9 to a new MAJOR key (G major).'' \\
\rowcolor{vGreen}
\textcolor{tGreen}{\Fxb} & Verdict (flip) & ``The piece modulates at bar 9 to a new MINOR key (A minor).'' \\
\rowcolor{vBlue}
\textcolor{tBlue}{\textbf{F2}}   & Key--value     & \texttt{\small bars9\_16\_key: G major} \\
\rowcolor{vBlue}
\textcolor{tBlue}{\textbf{F3}}   & Prose          & ``From bar 9 the harmony re-centers on G major, a major key (pivot chord D7).'' \\
\rowcolor{vRose}
\textcolor{tRose}{\textbf{F4}}   & Tool output    & \texttt{\small chord\_report: bar1=C, bar2=F, \ldots, bar9=D7, \ldots, bar16=G} \\
\rowcolor{vRose}
\textcolor{tRose}{\textbf{F5}}   & Output + noise & F4 $+$ \texttt{\small tempo\_bpm=104, first\_key=C major, rms\_db=-15.5} \\
\rowcolor{vRose}
\textcolor{tRose}{\textbf{F6}}   & JSON           & \texttt{\small \{"chords\_by\_bar": \{"1": "C", \ldots, "9": "D7", \ldots\}\}} \\
\bottomrule
\end{tabularx}
\end{table}

\noindent\textbf{Measurement formats.} For each item we express the same musical information in six forms that reflect how an MIR measurement may appear in a tool-augmented system (Table~\ref{tab:formats}). F1 and F3 state the conclusion; F2 gives the single value it follows from; F4 to F6 give the full tool output with increasing clutter (F5) and structure (F6). \Fx{} states the wrong conclusion in F1's form, to test whether a model follows the report even against the correct answer.

\noindent\textbf{Derivation ladder.} To order the formats by reasoning distance, we generate $72$ items per task in which the same fact is reported at four depths: d0 states the conclusion; d1 gives the fact one step away (for modulation, the home and new key); d2 gives the evidence for that fact (the pivot chord); d3 gives only its constituents (the chord's pitches). Items vary key and harmonic relation rather than transposition, so every report is distinct, and intervals for the ladder resample items.

\noindent\textbf{Real tool output and real recordings.} We rebuild all six formats for modulation and structure from a running librosa~\cite{librosa} analyzer on the same audio, so the report is a real tool's output rather than a template. Separately, on $60$ GuitarSet~\cite{guitarset} recordings with expert key annotations, we ask for the key from audio alone, with the annotation supplied as the tool output, and with a deliberately wrong annotation.

\noindent\textbf{Models and input conditions.} We test three open-weight models run locally, Qwen2.5-Omni~\cite{qwen}, Music Flamingo (MF)~\cite{musicflamingo}, and Audio Flamingo 3 (AF3)~\cite{af3}, and two API models, Gemini 2.5 Flash~\cite{gemini25} and Gemini 3.6 Flash~\cite{gemini36}. We run every format in two conditions: no-audio, where the model sees only the measurement, which isolates whether it can read the measurement on its own; and audio-crossed, where the model also gets the audio, which tests whether audio helps when reasoning from the measurement fails.

\noindent\textbf{Scoring.}
Each test item is a matched pair containing both alternatives of the task (e.g., one major-key modulation and one minor-key modulation). A pair is counted as correct only when the model answers \emph{both} members correctly. Because the two members have opposite correct answers, a model that always gives the same response is correct on at most one of them and scores $0\%$ pair accuracy; chance is $25\%$. We report 95\% percentile bootstrap intervals from 10,000 resamples over base compositions. We resample compositions rather than individual transpositions because transpositions of the same composition are not independent. We validated automatic keyword scoring against 100 manually scored responses, achieving $\kappa=0.90$ agreement.

\noindent\textbf{Knowledge probes.} To separate a missing fact from a failure to combine facts already present, we ask each fact the modulation ladder needs alone, with terse answer-only questions ($48$ items each): whether a named key is major or minor (needed at d1); which key a dominant seventh resolves to and whether a scale degree is major or minor (d2); which chord a set of pitches spells (d3). To check that a terse probe does not understate knowledge, the d2 rule is also asked with the rule stated, as multiple choice, and with reasoning permitted.

\noindent\textbf{Mechanistic probes.}
Behavioral failure alone cannot tell us whether Qwen2.5-Omni never
computes an answer or computes one it fails to use. We test this on two
conditions already in the behavioral evaluation: conclusion stated
(modulation d0, structure F1) and derivation required (modulation d2,
$72$ items; structure F4, $30$ items). Since the derivation-required
condition often declines to answer, we append a fixed continuation
(``The new key is'', ``The third section'') and read the residual
stream at the answer position.

A \emph{logit lens}~\cite{logitlens} projects that stream through the output head at
every layer, tracking the model's preference between the two candidate
answers: a conclusion that is present should be decodable before the
final layer. \emph{Activation steering}~\cite{actadd} tests whether the state that encodes the correct conclusion can rescue a failing derivation: for modulation, we inject the mean hidden-state difference between the two conditions, computed leave-one-out over items, into the derivation-required forward pass at one of seven layers and several strengths, and measure accuracy. As a control, we also state the required musical rule in the prompt (Fig.~\ref{fig:mech}).

\noindent\textbf{Two-call mitigation.} If a derivation is never
performed, forcing it into the transcript should help. We test this
with a two-call procedure: in the first call, the model states what the
measurement says and what it indicates; in the second call, it answers
using only its own first-call statement, with the original measurement
withheld. This differs from ordinary chain-of-thought
prompting~\cite{muse}, where the model can still fall back on the
original evidence.
\section{Results and Discussion}
\label{sec:results}

\finding{1}{Models follow stated conclusions, even when they are wrong.}
\noindent When the tool output states the conclusion (F1/F3), all models are
near-perfect. But when we deliberately flip that conclusion in \Fx{}, accuracy falls to $0\%$ for every model, the frontier models included (Table~\ref{tab:profile}, last column): each follows the incorrect statement, even when audio is available. Thus, the models read the tool output but do not reliably verify its conclusion.


\begin{table}[t]
\centering
\caption{Format profile, pair accuracy (\%), structure and modulation pooled ($n{=}60$ pairs, $12$ compositions). \textcolor{tGreen}{Verdict} $=$ F1/F3, \textcolor{tRose}{Measurement} $=$ F2/F4/F5/F6. $\Delta$ is verdict minus measurement, no audio [95\% CI] and with audio. \Fx{} is the flipped verdict (pair accuracy; $0$ means every pair followed the wrong statement).}
\label{tab:profile}
\footnotesize
\setlength{\tabcolsep}{3pt}
\renewcommand{\arraystretch}{1.05}
\begin{tabular}{@{}l cc cc c@{}}
\toprule
\rowcolor{hdr}
Model & \textcolor{tGreen}{Verdict} & \textcolor{tRose}{Meas.} & $\Delta$ no audio & $\Delta$ $+$aud. & \Fx{} \\
\midrule
Qwen2.5-Omni     & \cellcolor{vGreen}100 & \cellcolor{vRose}36 & 64 \Ci{56,72} & 48 & 0 \\
Music Flamingo   & \cellcolor{vGreen}100 & \cellcolor{vRose}35 & 65 \Ci{48,83} & 67 & 0 \\
Audio Flamingo 3 & \cellcolor{vGreen}100 & \cellcolor{vRose}35 & 65 \Ci{48,83} & 58 & 0 \\
Gemini 2.5 Flash & \cellcolor{vGreen}100 & \cellcolor{vRose}85 & 15 \Ci{9,21}  & 21 & 0 \\
Gemini 3.6 Flash & \cellcolor{vGreen}100 & \cellcolor{vRose}90 & 10 \Ci{4,17}  & 15 & 0 \\
\bottomrule
\end{tabular}
\end{table}

\finding{2}{Accuracy falls with the distance between tool output and answer; one reasoning step is enough.}

\noindent\textbf{The format profile.} The gap appears as soon as the tool provides measurements rather than the answer itself (Table~\ref{tab:profile}). In the no-audio condition, open-model accuracy drops from \textcolor{tRose}{\textbf{$100\% \to {\sim}35\%$}}, while the Gemini models show a smaller but consistent drop. Providing the relevant tool output is therefore not sufficient; performance depends on how much reasoning remains between that output and the answer. Modulation F2 reveals one confound: it gives the new key but not the home key, so the output alone does not establish that a modulation occurred. We therefore use a controlled derivation ladder in which every rung contains sufficient information, while only the number of reasoning steps changes.

\begin{figure}[t]
\centering
\includegraphics[width=\columnwidth]{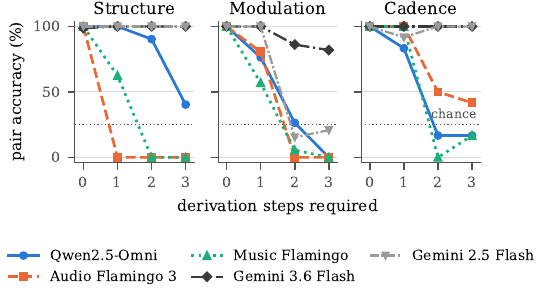}
\caption{Pair accuracy against derivation depth on three tasks, no audio, $n{=}72$ independent items each. Depth zero states the conclusion; each further step withholds one inference. Open models in color, frontier models in gray; dotted line is chance ($25\%$).}
\label{fig:depth}
\end{figure}

\noindent\textbf{The derivation ladder.} We fix the musical fact and vary how many reasoning steps separate the tool output from the answer (Fig.~\ref{fig:depth}, Table~\ref{tab:ladder}). On modulation, all three open models score $100\%$ when the report states the conclusion (d0). Replacing the conclusion with the new key, ``D major,'' leaves one step: recognizing that D major is a major key. Accuracy falls to as low as $57\%$. Giving the secondary dominant instead (d2) drops it to $0$--$26\%$, at or below chance. Giving only its pitches (d3) drops it to $0\%$. At every depth the question is the same, whether the new key is major or minor, and so is the answer: major (D major). Cadence and structure follow the same trend, though the step where models fail differs by task. The limitation appears before any long chain of reasoning. One elementary step between tool output and answer already costs up to $43$ points.
The frontier models split on the hardest chain: Gemini 2.5 Flash collapses at d2 like the open models ($15\%$), while Gemini 3.6 Flash, which reasons internally before answering, holds $86$ and $82\%$ at d2 and d3; both are at or near $100\%$ on every cadence and structure rung.

\finding{3}{The first-step failure is not a knowledge gap}
\label{sec:knowledge}
\noindent A natural explanation for the depth drop is missing music-theory knowledge. We test the required intermediate facts directly
(Table~\ref{tab:ladder}, right block). At d1, all three open models identify whether a named key is major or minor with 100\% accuracy, yet score as low as 57\% when that same fact must be used to answer the modulation question. Thus, even the first-step failure reflects difficulty applying available knowledge to the tool output, not simply missing knowledge. At deeper levels, both effects appear: some models lack the required theory, while others know the intermediate facts but still fail to combine them. The derivation failure therefore cannot be explained by knowledge gaps alone.

\begin{table}[t]
\centering
\caption{Modulation: derivation ladder beside the knowledge each depth needs. Color marks the reasoning required: \textcolor{tGreen}{none}, \textcolor{tBlue}{one step}, \textcolor{tRose}{two or more}. \emph{Audio, no report}: pair accuracy from the clip alone (chance $25$). \emph{Ladder}: pair accuracy (\%) without / with audio ($n{=}72$ items; one number when both agree). \emph{Knowledge}: the fact each depth needs, asked alone ($n{=}48$ per question for the open models, $37$ for the scale-degree question; $n{=}144$ for Gemini): d1, is a named key major or minor; d2, which key a dominant seventh resolves to / whether a scale degree is major or minor; d3, which chord a set of pitches spells. G2.5F/G3.6F: Gemini 2.5/3.6 Flash.}
\label{tab:ladder}
\scriptsize
\setlength{\tabcolsep}{1pt}
\renewcommand{\arraystretch}{1.05}
\begin{tabular}{@{}l c cccc ccc@{}}
\toprule
\rowcolor{hdr}
 & Audio, & \multicolumn{4}{c}{Ladder, without/with audio} & \multicolumn{3}{c}{Knowledge, asked alone} \\
\rowcolor{hdr}
Model & no report & \textcolor{tGreen}{d0} & \textcolor{tBlue}{d1} & \textcolor{tRose}{d2} & \textcolor{tRose}{d3} & \textcolor{tBlue}{d1} & \textcolor{tRose}{d2} & \textcolor{tRose}{d3} \\
\midrule
Qwen       & 17 & \cellcolor{vGreen}100 & \cellcolor{vBlue}76/89 & \cellcolor{vRose}26/25 & \cellcolor{vRose}0/0  & \cellcolor{vBlue}100 & \cellcolor{vRose}19/59  & \cellcolor{vRose}0  \\
AF3        & 0  & \cellcolor{vGreen}100 & \cellcolor{vBlue}81/8  & \cellcolor{vRose}0/0   & \cellcolor{vRose}0/0  & \cellcolor{vBlue}100 & \cellcolor{vRose}17/100 & \cellcolor{vRose}12 \\
MF         & 0  & \cellcolor{vGreen}100 & \cellcolor{vBlue}57/81 & \cellcolor{vRose}6/3   & \cellcolor{vRose}0/1  & \cellcolor{vBlue}100 & \cellcolor{vRose}85/59  & \cellcolor{vRose}19 \\
G2.5F & 17 & \cellcolor{vGreen}100 & \cellcolor{vBlue}100     & \cellcolor{vRose}15/43 & \cellcolor{vRose}21/32 & \cellcolor{vBlue}100 & \cellcolor{vRose}83/100 & \cellcolor{vRose}18 \\
G3.6F & 13 & \cellcolor{vGreen}100 & \cellcolor{vBlue}100     & \cellcolor{vRose}86/81 & \cellcolor{vRose}82/47 & \cellcolor{vBlue}100 & \cellcolor{vRose}85/100 & \cellcolor{vRose}23 \\
\bottomrule
\end{tabular}
\end{table}

\finding{4}{Audio does not rescue derivation, and the gap persists on real recordings.}

Providing the original audio does not recover failures at deeper derivation depths (Table~\ref{tab:ladder}, left). We then test whether the same pattern holds when the tool output comes from a running librosa analyzer \cite{librosa} rather than the template report, and whether it carries over to real GuitarSet recordings \cite{guitarset} (Table~\ref{tab:real}). When the analyzer is correct, conclusion-style outputs remain near-perfect, while measurement-style outputs reduce accuracy by 54--73 points across the three open models. When the analyzer gives an incorrect conclusion, the models follow it rather than correcting it from the audio, and on GuitarSet every model follows a wrong key annotation (33--100\% of the time) even though the open models identify the key from audio alone in 50--62\% of recordings and the Geminis in only 8--15\%. Thus, neither access to the audio nor real tool output nor real recordings removes the derivation gap.

\begin{table}[t]
\centering
\caption{Beyond the template report. \emph{Audio, no report}: accuracy from the recording alone (synthetic: pair accuracy, chance $25$; GuitarSet: open-ended key identification on $60$ recordings, chance about $4$). $\Delta$: verdict minus measurement formats, without / with audio, for the template report and for a running librosa analyzer on the same synthetic audio (analyzer correct on $56$ of $60$ pairs in the open-model runs, $54$ in the Gemini runs). \emph{Wrong verdict followed}: how often a wrong tool conclusion was adopted (synthetic \Fx{}; the analyzer's wrong pairs; a wrong key annotation on GuitarSet).}
\label{tab:real}
\scriptsize
\setlength{\tabcolsep}{1pt}
\renewcommand{\arraystretch}{1.05}
\begin{tabular}{@{}l cc cc ccc@{}}
\toprule
\rowcolor{hdr}
 & \multicolumn{2}{c}{Audio, no report} & \multicolumn{2}{c}{$\Delta$ verdict$-$meas.} & \multicolumn{3}{c}{Wrong verdict followed (\%)} \\
\rowcolor{hdr}
Model & Synth. & Guitar & Templ. & Librosa & \Fx{} & Librosa & Guitar \\
\midrule
Qwen       & 21 & 55 & \cellcolor{vRose}64/48 & \cellcolor{vRose}60/57 & \cellcolor{vGreen}100 & \cellcolor{vGreen}100 & \cellcolor{vGreen}63 \\
MF         & 0  & 62 & \cellcolor{vRose}65/67 & \cellcolor{vRose}73/52 & \cellcolor{vGreen}100 & \cellcolor{vGreen}100 & \cellcolor{vGreen}100 \\
AF3        & 0  & 50 & \cellcolor{vRose}65/58 & \cellcolor{vRose}54/61 & \cellcolor{vGreen}100 & \cellcolor{vGreen}100 & \cellcolor{vGreen}33 \\
\midrule
G2.5F & 17 & 8  & \cellcolor{vRose}15/21 & \cellcolor{vRose}33/23 & \cellcolor{vGreen}100 & \cellcolor{vGreen}100 & \cellcolor{vGreen}100 \\
G3.6F & 13 & 15 & \cellcolor{vRose}10/15 & \cellcolor{vRose}12/22 & \cellcolor{vGreen}100 & \cellcolor{vGreen}100 & \cellcolor{vGreen}100 \\
\bottomrule
\end{tabular}
\end{table}

\finding{5}{Derivation fails either because the conclusion is never formed or because it is lost.}

\begin{figure}[t]
\centering
\includegraphics[width=\columnwidth]{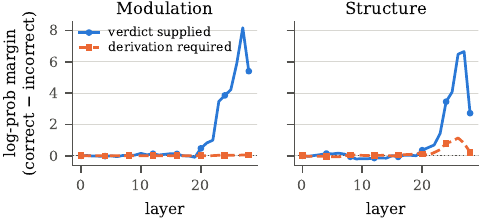}
\caption{Logit lens on Qwen2.5-Omni: log-probability margin between the correct and incorrect answer words decoded from each layer. With derivation required the margin never leaves zero for modulation (d2, $n{=}72$ items) but rises to $+0.9$ at layer $25$ for structure (F4, $n{=}30$) before falling at the output.}
\label{fig:mech}\label{fig:lens}
\end{figure}

\noindent The mechanistic probes reveal two distinct failure modes
(Fig.~\ref{fig:mech}). For modulation, the correct answer never becomes
reliably decodable at any layer when derivation is required. Activation
steering does not recover it (Qwen moves from $25\%$ to at most $29\%$ across seven layers), and stating the required rule in the prompt moves the open models nowhere (Qwen $26\to12$, AF3 $0\to0$, MF $6\to0$) while lifting the Geminis to $92$ and $99\%$. These three probes point to an \emph{absence} failure: the model does not appear to form the required conclusion.
Structure shows the opposite pattern. Although the final answer is wrong, the correct conclusion becomes decodable on 93\% of items at an intermediate layer before weakening toward the output. This is a \emph{loss} failure, the represented-but-unused pattern reported for speech prosody \cite{repignored}: useful information is formed internally but does not reliably reach the final prediction. Notably, structure is also the task on which Qwen remains robust to greater derivation depth, reaching 90\% at d2, linking stronger internal representation with greater behavioral robustness. If a conclusion is formed but lost, externalizing it may help; we test this next.

\finding{6}{Explicit intermediate reasoning helps selectively, but does not solve the problem.}

\begin{wraptable}{r}{0.5\columnwidth}
\vspace{-8pt}
\centering
\captionsetup{font=scriptsize,skip=2pt}
\scriptsize
\setlength{\tabcolsep}{2.5pt}
\renewcommand{\arraystretch}{1.0}
\begin{tabular}{@{}l ccc@{}}
\toprule
\rowcolor{hdr}
 & \textcolor{tBlue}{d1} & \textcolor{tRose}{d2} & \textcolor{tRose}{d3} \\
\midrule
Mod.   & \cellcolor{vGreen}$76{\to}79$ & \cellcolor{vRose}$26{\to}0$ & $0{\to}0$ \\
Cad.   & $83{\to}83$ & \cellcolor{vGreen}$17{\to}42$ & \cellcolor{vRose}$17{\to}8$ \\
Struct. & \cellcolor{vRose}$100{\to}62$ & \cellcolor{vRose}$90{\to}38$ & \cellcolor{vGreen}$40{\to}43$ \\
\bottomrule
\end{tabular}
\caption{Two-call mitigation, Qwen2.5-Omni: pair accuracy (\%), single $\to$ two-call, $n{=}72$. Column color: reasoning depth; cell color: direction of change.}
\label{tab:mitigation}
\vspace{-6pt}
\end{wraptable}
\noindent The two-call procedure (Sec.~\ref{sec:method}) helps only selectively (Table~\ref{tab:mitigation}). Cadence d2 improves from $17\%$ to $42\%$, and the failing structure full report on the battery from $0\%$ to $17\%$. But deeper
derivations are not rescued: modulation d3 remains at $0\%$, while cadence d3 falls from $17\%$ to $8\%$. The procedure can also destroy information that was previously usable: structure d1 drops from $100\%$ to $62\%$ and d2 from $90\%$ to $38\%$ when the intermediate summary omits the score the second call needs. Externalizing the intermediate step helps in a few cells, hurts in others, and does not extend the model's reasoning depth.

\section{Conclusion}
\label{sec:conclusion}
Given an MIR tool's output, current audio-language models copy a stated conclusion but rarely derive one, and the loss grows with every reasoning step left to them. Our results motivate evaluating tool-augmented audio systems not only by whether they retrieve the right information, but also by \textcolor{tBlue}{how much derivation they leave to the language model} and whether that derivation is performed reliably. This also suggests a training
direction: post-training should reward not only \textcolor{tBlue}{correct tool calling}, but also
\textcolor{tBlue}{correct tool reading}, that is, interpreting and reasoning over the returned output before answering. Derivation-depth examples and flipped-conclusion controls could provide supervision or reward signals for
this capability, while also testing whether models verify rather than simply follow tool outputs. Future work will extend these ideas to broader model families and musical systems, including non-Western traditions.

\section{Acknowledgment}

Rupali Rajesh was supported by the Post-Baccalaureate Research Fellowship
at the Wadhwani School of Data Science and AI, IIT Madras.

\section{Compliance with ethical standards}

All non-synthetic audio is from GuitarSet \cite{guitarset} (CC~BY~4.0); the
remaining stimuli are synthetic renders of programmatically generated scores
and carry no third-party rights. The human listening baseline was collected
from one consenting adult musician with no personal data recorded. No IRB
review was required. The authors declare no conflicts of interest.
Generative AI (Claude, Anthropic) assisted with generating the figures and with editing the manuscript for grammar and phrasing; the authors reviewed all content and
take full responsibility for its correctness.

\bibliographystyle{IEEEbib}
\bibliography{refs}

\end{document}